\newcommand{\eqn}[1]{\label{#1}}
\newcommand{\eq}[1]{\begin{equation}#1\end{equation}}
\newcommand{\eqnono}[1]{\begin{displaymath}#1\end{displaymath}}
\newcommand{\eqs}[1]{\begin{eqnarray}#1\end{eqnarray}}

\newcommand{\bra}[2]{{}_{#2}\big{<}#1\big{|}}
\newcommand{\ket}[2]{\big{|}#1\big{>}_{#2}}

\newcommand{\bin}[2]{\renewcommand{\arraystretch}{0.5}
		\renewcommand{\arraycolsep}{0pt}
		\left(\begin{array}{c}#1\\#2\end{array}\right)
		\renewcommand{\arraystretch}{1.0}
		\renewcommand{\arraycolsep}{3pt}}
\def\lb{\nonumber\\}
\def\dfn{=}

\newcommand{\Xp}[2]{\mbox{$X^{(#1,#2)}$}}
\newcommand{\Xt}[2]{\mbox{${\tilde X}^{(#1,#2)}$}}
\newcommand{\Xn}[2]{\mbox{$X_N^{(#1,#2)}$}}
\newcommand{\Xtn}[2]{\mbox{${\tilde X}_N^{(#1,#2)}$}}
\newcommand{\Xps}[1]{\mbox{$X^{(#1)}$}}
\newcommand{\Xts}[1]{\mbox{${\tilde X}^{(#1)}$}}
\newcommand{\Xns}[1]{\mbox{$X_N^{(#1)}$}}
\newcommand{\Xtns}[1]{\mbox{${\tilde X}_N^{(#1)}$}}
\newcommand{\Yp}[2]{\mbox{$Y^{(#1,#2)}$}}
\newcommand{\Yt}[2]{\mbox{${\tilde Y}^{(#1,#2)}$}}

\newcommand{\Ytn}[2]{\mbox{${\tilde Y}_N^{(#1,#2)}$}}
\newcommand{\Yps}[1]{\mbox{$Y^{(#1)}$}}
\newcommand{\Yts}[1]{\mbox{${\tilde Y}^{(#1)}$}}
\newcommand{\Yns}[1]{\mbox{$Y_N^{(#1)}$}}
\newcommand{\Ytns}[1]{\mbox{${\tilde Y}_N^{(#1)}$}}
\newcommand{\refbr}[1]{(\ref{#1})}

\documentstyle[12pt]{article}

\def\a{\alpha}

\def\d{\delta}
\def\e{\epsilon}

\def\l{\lambda}

\def\G{\Gamma}

\def\/{\over}
\def\*{\partial}
\def\|{\mid}

\def\linv{\l^{-1}}
\def\HH{{\cal H}}

\newcommand{\NP}[1]{Nucl. Phys.~{\bf #1}\ }
\newcommand{\PL}[1]{Phys. Lett.~{\bf #1}\ }

\newcommand{\NC}[1]{Nuovo Cimento~{\bf #1}\ }

\newcommand{\IJMP}[1]{Int. J. Mod. Phys.~{\bf #1} \ }
\newcommand{\MPL}[1]{Mod. Phys. Lett.~{\bf #1} \ }
\newcommand{\half}{1/2}
\begin{document}

\newtheorem{thm}{Theorem}
\newtheorem{prop}{Proposition}
\newtheorem{defn}{Definition}
\newtheorem{lemma}{Lemma}

\newpage
\pagenumbering{arabic}
\begin{flushleft}
G\"{o}teborg\\
ITP 94 - 9\\
hep-th/9405159\\
April 1994
\end{flushleft}
\begin{center}

{\huge The Twisted String Vertex Algorithm Applied to the $Z_2$-Twisted
Scalar String Four Vertex }\\[1cm]
{\large Niclas Engberg\footnote{tfene@fy.chalmers.se}\\
Bengt E. W. Nilsson\footnote{tfebn@fy.chalmers.se}\\
Anders Westerberg\footnote{tfeawg@fy.chalmers.se}\\
}
\vspace{1cm}
{\sl Institute of Theoretical Physics\\Chalmers University of Technology\\ and
G\"{o}teborg University\\
S-412 96 G\"{o}teborg, Sweden}
\end{center}
\vspace{0.5cm}
\begin{abstract}

Recently an algorithm was found by means of which one can calculate terms at
arbitrary oscillator level in the four-Ramond
 vertex obtained by sewing. Here we show that this algorithm is applicable also
to the case of ${\bf Z}_2$-twisted scalars
and derive the full propagator for scalars on the Riemann sphere with two
branch cuts. The relation to similar results
previously derived in the literature by other means is discussed briefly.
\end{abstract}
\newpage

\section{Introduction}

${\bf Z}_2$-twisted scalar fields, i.e. scalars with antiperiodic boundary
conditions on the circle, play an
important role in string theory. In the context of strings moving on orbifolds
(see e.g. ref.~\cite{DFMS87}) they arise in the analysis
of the twisted sectors which are of interest in e.g. phenomenological
applications of the theory. Scalars of this
kind also turn out to be of crucial importance in the vertex operator algebra
approach to the Monster \cite{FLM1,DGM1}.
For these reasons, it is worth while to develop further the techniques for
computing twisted string vertices with the goal of
making them applicable to vertices with an arbitrary
number of loops and external legs. In particular, understanding how to derive
such vertices by sewing will
be of importance in explicit constructions of string field theory
\cite{KZ92,Zwi93} on orbifolds and probably also for various generalizations
of vertex operator algebras of the type relevant for the Monster. In fact, in
the untwisted case, certain generalizations of vertex operator algebras are
known to be of importance in closed string field theory,
as discussed in e.g. ref.~\cite{Sta93} (for a brief introduction to vertex
operator algebras see ref.~\cite{Geb93}).

For untwisted fields sewing techniques are by now very well understood and can
rather easily be utilized to obtain
arbitrary untwisted vertices. Many of these results are reviewed in
ref.~\cite{DiV92}.
The technical reason for this success can be traced back to the appearance of a
group structure
generated by the infinite-dimensional matrices that arise in the exponent of
the sewn vertex. This group structure
makes it possible to rewrite all expressions in the vertex in the Schottky
representation of the resulting
Riemann surface (see refs.~\cite{DFHLPS1,ENS1} and references therein).
 Unfortunately, straightforward application of the techniques developed for
untwisted fields
 to the sewing of vertices with more than two external
twisted fields does not work. The failure is due to the fact that
the infinite-dimensional matrices that appear in these latter cases do not seem
to generate any kind of group
structure. Instead one has in the past resorted to other methods
\cite{CO}-\cite{DGM2} which, however,
 not until very recently \cite{ENS2}
were extended from the lowest lying modes to an arbitrary oscillator level. In
hindsight, knowing the answer for the
sewn vertex one may derive highly non-linear algebraic relations between
binomial coefficients which in some sense
play the role of the group structure in the untwisted case. For the lowest
lying Ramond modes the binomial coefficients
are Catalan numbers and the algebraic relations among them are derived in
refs.~\cite{NS92,ENS2}.

In order to describe the sewing procedure more explicitly we can divide it into
two
major steps. The first one is basically trivial
and consists of evaluating an infinite set of gaussian integrals. It can be
carried out in
all cases, untwisted as well as twisted,
as long as the fields are expandable in Laurent series and the exponents of
the vertices to be sewn are at most bilinear
in these fields. The second step is the non-trivial one and amounts to
finding a way to rewrite the answer obtained in
the first step in terms of functions defined directly on the Riemann surface
produced by the sewing\footnote{This step is
discussed in great detail for the untwisted bosonic string in
refs.~\cite{LPP1,LPP2}}. As already alluded to above,
by using the Schottky representation of tau-functions etc. on Riemann surfaces
without branch cuts, the second step can be carried
out in all cases involving at most two twisted external legs, with or without
background charges (see ref.~\cite{DiV92} and references therein).
 For more general twisted vertices,
closed expressions have been derived in refs.~\cite{CO}-\cite{CH2} but for the
lowest modes only and only in the case of the four vertex.
 For twisted scalars some additional next to lowest
level results are presented in ref.~\cite{DGM2}. However, a
 systematic way to obtain closed formulae at any oscillator level
was found only recently \cite{ENS2} in the context of ${\bf Z}_2$ fermions,
i.e. Ramond fields. The form of the propagator obtained by
means of this twisted string vertex algorithm was later derived analytically
in ref.~\cite{NS93}.

The case of twisted scalars is in this context technically more involved than
the case of twisted fermions. Nevertheless,
we will in this paper demonstrate that the algorithm just mentioned can be
carried over to the vertex for four twisted scalars
and be successfully used to obtain the explicit form of the propagator on the
twice cut Riemann sphere. (A vertex for an arbitrary number of external twisted
fermions was proposed
in refs.~\cite{DHMR,DMR} based on path integral arguments. It is unclear if
these arguments can be derived from the operator
methods used in this paper.) As we will see later,
the dependence on the modulus $\l$ (the end points of the two cuts are located
at $(0,-\l)$ and $(-{\linv},\infty)$, respectively,
 with $0< \l <1$) is more intricate in the scalar case. The propagator
 for the left-movers derived in section three below depends on $\l$ through
the complete elliptic integrals of the first and second kind, $K(\l)$ and
$E(\l)$, which do not appear at all for
 the case of twisted fermions \cite{ENS2}.  Our result for twisted scalars
turns out to coincide with that of Dixon et al. \cite{DFMS87},
but only in the limit of vanishing $\l$. On the other hand, for the modes
where a direct comparison of our results with those of Dolan et al.~\cite{DGM2}
is possible, complete
agreement is established.

This paper is organized as follows. In section two we present in some detail
the three vertex for one untwisted and
two twisted external scalar legs in its dual form \cite{ENS1}, and give the
result of the first step
 in the sewing procedure explained above.
The algorithm of ref.~\cite{ENS2} is then applied in section three and a large
number of terms in the propagator are computed explicitly.
 These
terms are then used in section four to deduce the closed form of the propagator
which is easily seen to produce, after the evaluation
of two analytic integrals, each one of the terms quoted in section three. Some
concluding remarks are collected in section five.

\section{Sewing of the String Vertex for Four Twisted Scalars}
\label{Sewsec}

In this section we briefly review how the vertex describing the scattering of
four
external (${\bf Z}_2$) twisted scalar fields can be constructed by sewing
together two dual twisted string vertices.

An ordinary (untwisted) scalar field $\hat{x}(z)$ has periodic boundary
conditions and a
mode expansion\footnote{All space-time indices will be suppressed throughout
this paper. Also, all fields considered are chiral.}
\eq{\hat{x}(z)=\hat{q}-i\hat{p}\log z + i\sum_{n\neq 0}{\hat{\a}_n \/n}z^{-n},\
\ \ \
[\hat{\a}_n,\hat{\a}_m]=n\d_{n+m,0},}
corresponding to the usual commutator between the annihilation and creation
parts:
\eq{[\hat{x}^{(+)}(z),\hat{x}^{(-)}(w)]=-\log (z-w), \ \ \ \ |z|>|w|.}
For a (${\bf Z}_2$) twisted scalar field $\hat{x}_{tw}$ the boundary conditions
are
antiperiodic, which leads to a mode expansion of the form
\eq{\hat{x}_{tw}(z)=i\sum_{r\in {\bf Z}+\half}{\hat{c}_r \/ r}z^{-r},\ \ \ \
[\hat{c}_r ,\hat{c}_s ]=r\d_{r+s,0}.}
In this case the commutator for the annihilation and creation parts is
\eq{[\hat{x}_{tw}^{(+)}(z),\hat{x}_{tw}^{(-)}(w)]=\log \left(
{\sqrt{z}+\sqrt{w} \/ \sqrt{z}-\sqrt{w}}\right),\ \ \ \ |z|>|w|.}

The starting point of this paper is the dual Reggeon vertex for twisted
scalars. A dual vertex \cite{NTWH,NT1} is a vertex that is dual
(in the sense of dual models) to the OPE of two ordinary three vertices.
The dual vertices are convenient
to use when dealing with twisted fields,
because twisted fields must always appear in pairs and the dual vertex
provides a very natural pairing of the external twisted states\footnote{Dual
vertices can
sometimes greatly simplify calculations also for untwisted fields. This is
seen e.g. in the loop calculations of refs.~\cite{ENS1,E1}.}.
The dual vertex describing the emission or absorption of two external
twisted scalars was constructed in ref.~\cite{ENS1} by sewing
together two ordinary three vertices. The result can be written in the form
\begin{eqnarray}
\lefteqn{\hat{W}^{(tw)}(V_{i}) =} \lb
& & \bra{q=0}{ord}\int dk_{i}:\exp{\left(
{1\/2}k_{i}^{2}\log{(V_{i}\G)'(0)}+ik_i\{(\hat{x}_{tw}+i\hat{x}_{ord})
(V_{i}^{-1}(\infty))-\hat{x}_{aux}(V_{i}(\infty))\}\right)}\lb
& & \times\exp{\left(-\oint_{0,\infty}dz\hat{x}_{aux}(V_{i}(z))\partial
(\hat{x}_{tw}+i\hat{x}_{ord})(z)\right)}:\ket{p=0}{ord}
\end{eqnarray}\label{dualvertex}
where the normal ordering in the twisted sector $\HH_{tw}$ of the Hilbert space
has been implemented by using an untwisted
normal ordering field $\hat{x}_{ord}$ \cite{ENS1}. Such a normal ordering
field,
to our knowledge, was used for the first time in ref.~\cite{CO} but only in
defining the Ramond three vertex.
The normal ordering dots in eq.~\refbr{dualvertex} refer to the (untwisted)
auxiliary field $\hat{x}_{aux}$.
Furthermore, $V_{i}(z)$ is a projective transformation defining the emission
points, one at  $V_{i}(0)$ and the other at $V_{i}(\infty)$.

The four-Reggeon vertex is the result of multiplying two dual vertices and
computing a vacuum correlation in the auxiliary Hilbert space $\HH_{aux}$:
\eq{\hat{W}_{4}^{(tw)}(V_{1},V_{2})=\bra{p=0}{aux}\hat{W}^{(tw)}(V_{1})
\hat{W}^{(tw)}(V_{2})\ket{p=0}{aux}.}
Due to the projective invariance of the $\HH_{aux}$ vacuum we can choose the
transformations $V_{1}$ and $V_{2}$ arbitrarily as long as we keep only one
moduli parameter $\l$. A convenient choice is
\eq{V_{1}^{-1}(z)={1\/ z+{\linv}},\ \ \ \ V_{2}^{-1}(z)=z^{-1}+{\linv},}
corresponding to the emission points $V_{1}(0)=\infty,\, V_{1}(\infty )=
-{\linv},\, V_{2}(0)=-\l$ and $V_{2}(\infty)=0$.
However, when eliminating the normal ordering fields, using this choice of
transformations, we find that in some of
the resulting terms the end points of the cuts of different factors coincide.
We therefore introduce a point splitting parameter $\e$ according to
\eq{V_{1}^{-1}(z)={1-\e z\/ z+{\linv}},\ \ \ \
V_{2}^{-1}(z)={z\linv+1\/ z+\e}.}
It is then straightforward to show that after the sewing has been carried out
all terms singular in $\e$ cancel so that in the end we can take the limit $\e
\rightarrow 0$
without problems.
After elimination of the normal ordering fields the four-Reggeon vertex takes
the form
\eqs{
\lefteqn{\hat{W}_{4}^{(tw)}=\int dk:\exp{\left(-k^{2}D_{00}({\linv})+
k[\hat{c}_{r}^{(1)}V_0^{(-r)}({\linv})+
\hat{c}_{r}^{(2)}V_0^{(r)}({\linv})]\right)}}\lb
& & \times\bra{0}{aux \,osc.}\exp{\left(-{1\/2} \hat{\a}_{n}\hat{\a}_{m}
D_{mn}({\linv})-k\hat{\a}_{m}D_{0m}({\linv})+
\hat{c}_{r}^{(1)}\hat{\a}_{m}
V_m^{(-r)}({\linv})\right)}\lb
& & \times\exp{\left(-{1\/2} \hat{\a}_{-m}\hat{\a}_{-n}
D_{mn}({\linv})-k\hat{\a}_{-m}D_{0m}({\linv})+
\hat{c}_{r}^{(2)}\hat{\a}_{-m}
V_m^{(r)}({\linv})\right)}:\ket{0}{aux \,osc.},\label{matrixform}}
where we have introduced the infinite-dimensional matrix
$D({\linv})$~\cite{SchJ1} with components
\eq{D_{00}({\linv})=\log{4\/ \l},\ \ \ \
D_{mn}({\linv})=-{1\/m+n}\bin{-1/2}{m}\bin{-1/2}{n}\l^{m+n},\ \ \ m,n\in {\bf
N},\ (m,n)\neq(0,0)}
and the infinite-dimensional vectors $V^{(r)}\ (r\in{\bf Z}+\half)$ defined by
\eq{V_m^{(r)}({\linv})={1\/ r}\bin{-r}{m}\l^{r+m},\ \ \ \ m\in {\bf Z}_+.}
(We use the convention that repeated indices $m,n$ are summed over ${\bf Z_+}$
and $r,s$ over ${\bf Z}+\half$.)
The normal ordering in eq.~\refbr{matrixform} now refers only to the twisted
oscillators. Because of the terms bilinear in the $\hat{x}_{aux}$ oscillators
$\hat{\a}$ we cannot use the Baker--Hausdorff formula to evaluate the
remaining $\HH_{aux}$ correlation. Instead we insert a
coherent state completeness relation and perform the resulting
infinite-dimensional integral.
In $d$ spacetime dimensions the result of the integration reads
\eq{\hat{W}_{4}^{(tw)}= \left(\det{G}\right)^{-{d\/ 2}}\int
dk\exp{(A+{1\/2}b^{T}G^{-1}Pb)},}
where
\eqs{G_{mn}=\left(\begin{array}{cc} {\bf 1} & \sqrt{mn}D_{mn}({\linv}) \\
 \sqrt{mn}D_{mn}({\linv}) & {\bf 1} \end{array}\right),\;\;\;\;
P=\left(\begin{array}{cc} 0 & {\bf 1} \\ {\bf 1} & 0 \end{array} \right), \lb
A=-k^{2}D_{00}({\linv})+k\hat{c}_{r}^{(1)}V_0^{(-r)}({\linv})+
k\hat{c}_{r}^{(2)}V_0^{(r)}({\linv}),\;\;\;\;
b_m=\left(\begin{array}{c}\sqrt{m}B_{m}\\ \sqrt{m}B_{-m}\end{array}\right), \lb
B_{m}=-kD_{0m}({\linv})+\hat{c}_{r}^{(1)}V_m^{(-r)}({\linv}),\;\;\;\;
B_{-m}=-kD_{0m}({\linv})+\hat{c}_{r}^{(2)}V_m^{(r)}({\linv}).
}
Introducing the matrices
\eq{S_{mn}({\linv})={m\/ m+n}\bin{-\half}{m}\bin{-\half}{n}\l^{m+n},\ \ \ \
N_{mn}=m\d_{m,n},\ \ \ \ m,n\in{\bf Z}_+ \eqn{sndef}}
we find that, after inverting $G$, the vertex becomes
\eqs{
\lefteqn{\hat{W}_{4}^{(tw)}=\left( \det{G}\right)^{-{d\/ 2}}\int dk
:\exp{\left( A +
	\left\{k^{2}D_{0m}\left({1\/ 1-S}N\right)_{mn}D_{n0}\; \right.\right.}} \lb
& & +k\left[
	\hat{c}_{r}^{(1)}V_m^{(-r)}\left( {1\/ 1-S}N \right)_{mn}D_{n0}-
	\hat{c}_{r}^{(2)}V_m^{(r)}\left( {1\/ 1-S}N \right)_{mn}D_{n0} \right] \lb
& & +{1\/2}\left[
	\hat{c}_{r}^{(1)}\hat{c}_{s}^{(1)}V_m^{(-r)}\left({S\/
1-S^{2}}N\right)_{mn}V_n^{(-s)}+
	2\hat{c}_{r}^{(1)}\hat{c}_{s}^{(2)}V_m^{(-r)}\left({1\/
1-S^{2}}N\right)_{mn}V_n^{(s)}
		\right. \lb
& &+\left.\left.\left.\hat{c}_{r}^{(2)}\hat{c}_{s}^{(2)}V_m^{(r)}\left({S\/
1-S^{2}}N\right)_{mn}V_n^{(s)}
		\right]\right\}\right):,
	\label{sewedvertex}
}
where \cite{SW2}
\eq{\det{G(\l)}={2\/ \pi}(1-\l^{2})^{1\/ 4}K(\l),}
with $K$ denoting the complete elliptic integral of the first kind.

\section{Application of the twisted string vertex algorithm}

To evaluate the terms in the exponent of $\hat{W}_{4}^{(tw)}$ we use a modified
version of the
algorithm found in ref.~\cite{ENS2}. Thus, we start by defining the
infinite-dimensional vectors
\eqs{
\xi^{(r)}=(1+S)^{-1}v^{(r)}, &    & \eta^{(r)}=(1-S)^{-1}v^{(r)},\lb
\tilde{\xi}^{(r)}=(1+S^{T})^{-1}v^{(r)}, &    &
\tilde{\eta}^{(r)}=(1-S^{T})^{-1}v^{(r)},
\eqn{rvecdefs}
}
where $r\in{\bf Z}+\half$ and $v^{(r)}$ is given by
\eq{
v_{m}^{(r)}={1\/ \sqrt{2}}\bin{-r}{m}\l^{r+m},\ \ m\in{\bf Z}_+.\label{vdef}
}
It turns out to be convenient to define also
\eqs{
\xi =\l^{(-\half)}\xi^{(\half)}=(1+S)^{-1}v, &    &
\eta=\l^{(-\half)}\eta^{(\half)}=(1-S)^{-1}v, \lb
\tilde{\xi}=\l^{(-\half)}\tilde{\xi}^{(\half)}=(1+S^T)^{-1}v, &    &
\tilde{\eta}=\l^{(-\half)}\tilde{\eta}^{(\half)}=(1-S^T)^{-1}v,
\eqn{vecdefs}
}
with $v=\l^{-\half}v^{(\half)}$.

Making use of the matrix identity $A^{-1}(A+B)B^{-1}=A^{-1}+B^{-1}$ and the
fact that
$N^{-1}SN=S^T$, we obtain the relations
\eqs{
V_m^{(r)}\left({1\/ 1-S^{2}}N\right)_{mn}V_n^{(s)}&=&{1\/ rs}\left(
\Ytn{r}{s}+\Xtn{r}{s}\right),\\
V_m^{(r)}\left({S\/ 1-S^{2}}N\right)_{mn}V_n^{(s)}&=&
{1\/ rs}\left(\Ytn{r}{s}-\Xtn{r}{s}\right),\\
V_m^{(r)}\left({1\/ 1-S}N\right)_{mn}D_{n0}&=&\mbox{}-{2\/r}\Yps{r},}
where
\eq{
\Xtn{r}{s} \dfn v^{(r)T}N\tilde{\xi}^{(s)},\ \ \ \Ytn{r}{s} \dfn
v^{(r)T}N\tilde{\eta}^{(s)}, \ \ \
\Yps{r} \dfn v^{(r)T}\eta.\label{osccoeffs}
}
All oscillator coefficients in eq.~\refbr{sewedvertex} can thus be expressed in
terms of \Xtn{r}{s},
\Ytn{r}{s} and \Yps{r}, the evaluation of which is the subject of the remainder
of this section.

To this end we will find use for the quantities
\eqs{
\Xp{r}{s}\dfn v^{(r)T}\xi^{(s)},\ \ \ \Xn{r}{s}\dfn v^{(r)T}N\xi^{(s)}, \ \ \
\Xt{r}{s}\dfn v^{(r)T}\tilde{\xi}, \lb
X\dfn v^T\xi,\ \ \ \Xps{r}\dfn v^{(r)T}\xi,\ \ \ \Xns{r}\dfn v^{(r)T}N\xi,\ \ \
\Xts{r}\dfn v^{(r)T}\tilde{\xi},
}
defined in analogy with~\refbr{osccoeffs}, as well as for the corresponding
$\eta$-dependent
quantities which we label $Y$ instead of $X$. It follows from the
definitions~\refbr{vecdefs} that
$\Xps{\half}=\l^{\half}X$, $\Xp{r}{\half}=\l^{\half}\Xps{r}$, etc. We will also
need the following
readily derived transformation properties under index permutation:
\eq{
\Xp{r}{s}=\Xt{s}{r},\ \ \ \Xtn{r}{s}=\Xtn{s}{r}. \label{indexperm}
}
Analogous identities are valid for \Yp{r}{s}, \Yt{r}{s} and \Ytn{r}{s}.

The vectors $v$, $\xi$ and $\eta$ were introduced in ref.~\cite{SW2} where it
was also  shown that $\xi$
and $\eta$ are related according to
\eq{ \xi_{n}={1\/ ng(\l )}\l {\partial \/ \partial \l}\eta_{n}
\label{xietarel}}
and satisfy the differential equations
\eqs{\l {\partial \/ \partial
\l}\left( g(\l )\l {\partial \/ \partial \l} \xi  \right) &=&N^{2}\xi g(\l ),
\label{xieq} \\ \l
{\partial \/ \partial \l}\left( {1\/ g(\l )}\l  {\partial \/ \partial \l} \eta
\right) &=&{N^{2}\eta \/ g(\l ),\label{etaeq}}
}
where $g$ is defined by
\eq{g(\l )={1+2Y \/ 1-2X}={1+2v^{T}\eta \/ 1-2v^{T}\xi}=\left( {2\/ \pi}
\sqrt{1-\l^{2}}K(\l )\right)^{-2}.}
The solution of eq.~\refbr{etaeq} is~\cite{SW2}
\eq{\eta_{n}={\pi \/ 2\sqrt{2}K(\l)}\bin{-1/2}{n}\l^{n} \,_{2}F_{1}(\half
,n+\half , n+1,\l^{2}),}
which, by using the integral representation of the hypergeometric function
$_2 F_1$,
can be written as
\eq{\eta_{n}={i\pi \/ 2\sqrt{2}K(\l)}\oint_{C}{(-\l t)^{n}dt \/
(t(1-t)(1-\l^{2}t))^{\half} },\label{etaintrep}}
with the contour $C$ enclosing the cut between $0$ and $1$.
The corresponding integral representation for $\xi$ then follows from
eq.~\refbr{xietarel}:
\eqs{\xi_{n} &=& {i\sqrt{2}\/ \pi}\left[ (1-\l^{2})K(\l)-{E(\l)\/ n}\right]
\oint_{C}{(-\l t)^{n}dt\/ (t(1-t)(1-\l^{2}t))^{1\/ 2}} \lb
 & & +{i\sqrt{2}\/\pi}(1-\l^{2}){K(\l) \/ n}\oint_{C}{(-\l t)^{n}dt\/
(t(1-t))^{1\/ 2}
(1-\l^{2}t)^{3\/ 2}}.\label{xiintrep}}
Here $E$ denotes the complete elliptic integral of the second kind.

Using the integral representations for $\eta_{n}$ and $\xi_{n}$ we can then
calculate \Yps{r} as
well as \Xns{r} and \Yns{r} at arbitrary level. When reversing the order of
summation and
integration, the cuts in the integrands either cancel or become poles, leaving
only
some quite straightforward integrals. Some explicit results for \Yps{r} are
shown in
table~\ref{vTetatabell}.
\begin{table}
\eqnono{\begin{array}{|c|c|}\hline
r & \Yps{r}\dfn v^{(r)T}\eta \\ \hline
{1\/ 2} & \l^{{1\/ 2}}\left( {\pi \/ 4K}(1-\l^{2})^{-{1\/ 2}}-{1\/ 2}\right) \\
 \hline
-{1\/ 2} & \l^{-{1\/ 2}}\left( {\pi \/ 4K}-{1\/ 2}\right) \\ \hline
{3\/ 2} & \l^{3\/ 2}\left( {\pi \/ 4K}(1-{\l^{2}\/ 2})(1-\l^{2})^{-{3\/ 2}}-
{1\/ 2}\right) \\ \hline
-{3\/ 2} & \l^{-{3\/ 2}}\left( {\pi \/ 4K}(1-{\l^{2}\/ 2})-{1\/ 2}\right) \\
\hline \end{array}}
\caption{Some of the quantities that can be calculated directly from the
integral representation of $\eta$.} \label{vTetatabell} \end{table}

Contrary to the case of \Yps{r}, the quantities \Xtn{r}{s} and \Ytn{r}{s} must
be calculated recursively.
In order to see how this is accomplished, we first note that the vectors
$v^{(r)}$ satisfy the
recursion relation
\eq{ v^{(r)}=\l \left( 1+{N\/ r-1}\right)v^{(r-1)}\label{vrec}}
as a consequence of the definitions and the identity
$\bin{m-1}{n}=(1-{n\/m})\bin{m}{n}$.
Using the definitions~\refbr{rvecdefs} we then immediately obtain the
corresponding recursion
relations for $\xi^{(r)}$and $\eta^{(r)}$:
\eq{ \eta^{(r)}=\l \eta^{(r-1)}+{\l N\/ r-1}\tilde{\eta}^{(r-1)},\ \ \ \
\xi^{(r)}=\l \xi^{(r-1)}+{\l N\/ r-1}\tilde{\xi}^{(r-1)}.\label{xietarec}}

Furthermore, the matrix identity $A^{-1}(A-B)B^{-1}=B^{-1}-A^{-1}$ and the fact
that
$S+S^{T}=2vv^{T}$ give the relations
\eq{ \tilde{\xi}={\eta \/ 1+2v^{T}\eta}, \ \ \ \
\tilde{\eta}={\xi \/ 1-2v^{T}\xi}. \label{xietatilde}}
Similarly we find that
\eqs{\tilde{\eta}^{(s)}=\xi^{(s)}+2\tilde{\eta}v^{T}\xi^{(s)},\ \ \ \
\tilde{\xi}^{(s)}=\eta^{(s)}-2\tilde{\xi}v^{T}\eta^{(s)}.\label{rxietatilde}}
When multiplied by $v^T$ from the left, both equations in~\refbr{xietatilde}
yield the identity
\eq{(1-2X)(1+2Y)=1,\label{XYrel}}
using the additional facts that
$v^T\tilde{\xi}=\l^{-1}\Xt{\half}{\half}=\l^{-1}\Xp{\half}{\half}=X$ and
analogously $v^T\tilde{\eta}=Y$.

By transposing eq.~\refbr{vrec} we can derive recursion formulas for \Xps{r}
and \Yns{r}:
\eqs{
\Xps{r}=\l\left({1\/{r-1}}\Xns{r-1}+\Xps{r-1}\right)=
{1\/{\l}}\Xps{r+1}-{1\/r}\Xns{r},\label{Xpsrec}\\
\Yns{r}=r\left({1\/{\l}}\Yps{r+1}-\Yps{r}\right)\label{Ynsrec},
}
where in~\refbr{Xpsrec} $\Xps{\half}=\l^{\half}X$ is known from~\refbr{XYrel}.
Eq.~\refbr{Ynsrec} is
not really needed since \Yns{r} can be obtained from the integral
representation of $\eta$, but it
nevertheless gives a more efficient way to calculate these quantities.
The identities \refbr{xietatilde} then relate \Xts{r}, \Yts{r}, \Xtns{r} and
\Ytns{r} directly to
quantities that we already know how to calculate:
\eqs{
\Xts{r}={1\/{1+2Y}}\Yps{r}, &    & \Yts{r}={1\/{1-2X}}\Xps{r}, \lb
\Xtns{r}={1\/{1+2Y}}\Yns{r}, &    & \Ytns{r}={1\/{1-2X}}\Xns{r}.
}

We are now in a position to derive the recursion relations that
allow us to calculate \Xtn{r}{s} and \Ytn{r}{s} as long as $r$ and $s$ are not
both negative;
thus, using eqs.~\refbr{indexperm}, \refbr{xietarec}, \refbr{xietatilde},
\refbr{rxietatilde} and
\refbr{XYrel} we arrive at the equations
\eqs{
\Xtn{r}{s} &=&
r\left[\Yp{r+1}{s-1}+{1\/{s-1}}\Ytn{r+1}{s-1}-
\l\Yp{r}{s-1}-{\l\/{s-1}}\Ytn{r}{s-1}
\right]-2\Yns{r}\Xps{s},\\
\Xp{r}{s} &=& \l\left[\Xp{r}{s-1}+{1\/{s-1}}\Xtn{r}{s-1}\right], \\
\Ytn{r}{s} &=&
r\left[\Xp{r+1}{s-1}+{1\/{s-1}}\Xtn{r+1}{s-1}-
\l\Xp{r}{s-1}-{\l\/{s-1}}\Xtn{r}{s-1}
\right]+2\Xns{r}\Yps{s},\\
\Yp{r}{s} &=& \l\left[\Yp{r}{s-1}+{1\/{s-1}}\Ytn{r}{s-1}\right].
}
The importance of \Xps{r}, \Xts{r}, \Yts{r}, \Xns{r}, \Yns{r}, \Xtns{r} and
\Ytns{r} comes from the
fact that they
supply the initial values needed to solve these equations recursively for
\Xtn{r}{s} and \Ytn{r}{s}.
We have, for instance, $\Xp{\half}{r}=\Xt{r}{\half}=\l^{\half}\Xts{r}$.

If $r$ and $s$ are both negative the above recursion relations cannot be
applied.
Instead we take the derivative of
\Xtn{r}{s} and \Ytn{r}{s} with respect to  $\l$.
For \Xtn{r}{s} this gives the differential equation
\eq{{\partial \/ \partial \l}\Xtn{r}{s}={r\/ \l^{2}}
\Xtn{r+1}{s}-{2\/ \l}\Xtns{r}\Xtns{s}+{s\/ \l^{2}}\Xtn{r}{s+1},}
with a completely analogous expression for \Ytn{r}{s}.
Now, since no terms on the right hand side have indices lower than $r$ or $s$,
the right hand side is expressed in terms of known quantities and we get
\Xtn{r}{s} by integrating with respect to $\l$. The value of the integration
constant is determined by examining the $\l^0$ term in  the Taylor
expansion for $\Xtn{r}{s}\dfn v^{(r)T}N\tilde{\xi}^{(s)}$.
We are thus able to calculate the quantities occurring in the exponent
to any given order in $r$ and $s$, i.e. for any choice of the external states.
Some of the relevant terms corresponding to the lowest levels
are collected in table \ref{tildetabell}.
\begin{table}
\eqnono{\begin{array}{|c|c|c|} \hline
r,s & \Xtn{r}{s} & \Ytn{r}{s} \\ \hline
{1\/ 2},{1\/ 2} & {\l^{3}\/ 8}(1-\l^{2})^{-1} & \l \left( {1\/ 8}+
{3\/ 8}(1-\l^{2})^{-1}-{1\/ 2}{E\/ K}(1-\l^{2})^{-1}\right) \\ \hline
-{1\/ 2},{1\/ 2} & {1\/ 4}\left( (1-\l^{2})^{1\/ 2}-1\right) &
{1\/ 4}\left( 1+(1-\l^{2})^{1\/ 2}\right)-{1\/ 2}{E\/ K}(1-\l^{2})^{-{1\/ 2}}
\\ \hline
-{1\/ 2},-{1\/ 2} & {\l \/ 8} & -{\l \/ 8}+{1\/ 2\l}\left( 1-{E\/ K}\right)
\\ \hline
{3 \/ 2},{1\/ 2} & {3\l^{4}\/ 8}(1-\l^{2})^{-2}(1-{\l^{2}\/ 4}) &
\l^{2}(1-\l^{2})^{-2}\left( {1\/ 2}-{\l^{2}\/ 8}+{3\l^{4}\/ 32}-{1\/ 2}{E\/ K}
(1-{\l^{2}\/ 2})\right) \\ \hline
-{3\/ 2},{1\/ 2} & -{3\l \/ 8}(1-\l^{2})^{1\/ 2} & -{3\l \/ 8}
(1-\l^{2})^{1\/ 2}-{1\/ 2\l}{E\/ K}(1-\l^{2})^{-{1\/2}}(1-{\l^{2}\/ 2})+
{1\/ 2\l}(1-\l^{2})^{1\/ 2} \\ \hline
-{1\/ 2},{3\/ 2} & -{3\l^{3}\/ 8}(1-\l^{2})^{-{1\/ 2}} &
{\l \/ 2}(1-\l^{2})^{-{3\/ 2}}\left( 1-{7\l^{2}\/4}+{3\l^{4}\/ 4}-
{E\/ K}(1-{\l^{2}\/ 2})\right) \\ \hline
-{3\/ 2},-{1\/ 2} & {3\/ 8}-{3\l^{2}\/ 32} & {3\l^{2}\/ 32}+{1\/ 4}{E\/ K}+
{1\/ 2\l^{2}}-{1\/ 2\l^{2}}{E\/ K}-{1\/ 8} \\ \hline
{3\/ 2},{3\/ 2} & {9\l^{5}\/8}(1-\l^{2})^{-3}(1+{\l^{4}\/ 12}) &
{\l^{3}\/ 2}(1-\l^{2})^{-3}\left( 1+{3\l^{2}\/ 4}+{5\l^{4}\/4}-{3\l^{6}\/ 16}-
{E\/ K}(1-\l^{2}+{\l^{4}\/ 4})\right) \\ \hline
{3\/ 2},-{3\/ 2} & -{3\/ 4}+{3\/ 4}(1-\l^{2})^{-{1\/2}}(1-2\l^{2}+
{7\l^{4}\/ 4}) & {3\/ 4}-{(1-\l^{2})^{-{3\/ 2}}\/ 4}\left( 1+3\l^{2}-
{37\l^{4}\/ 4}+{21\l^{6}\/ 4}+2{E\/ K}(1-\l^{2}+{\l^{4}\/ 4})\right)
 \\ \hline
-{3\/ 2},-{3\/ 2} & {9\/ 8\l}+{3\l^{3}\/ 32} & {5\l \/ 8}+{\l^{-3}\/ 2}+
{3\/ 8\l}-{3\l^{3}\/ 32}-{\l \/ 8}{E\/ K}-{\l^{-3}\/ 2}{E\/ K}+
{1\/ 2\l}{E\/ K}\\ \hline \end{array}}
\caption{The quantities occurring in the exponent for the lowest levels.}
\label{tildetabell}
\end{table}

Finally, the term in \refbr{sewedvertex} quadratic in the momenta $k$ can be
evaluated
by first taking the derivative
\eq{{\partial \/ \partial \l}\left(
D_{0m}\left({1\/ 1-S}N\right)_{mn}D_{n0}\right)=
2\linv\left(Y+X g(\l)\right)=-{\linv}+{\pi^{2}\/ 4\l
(1-\l^{2})K(\l)^{2}}}
and then integrating to obtain
\eq{D_{0m}\left({1\/ 1-S}N\right)_{mn}D_{n0}=\log{4\/ \l}-{\pi \/ 2}
{K(\sqrt{1-\l^{2}})\/ K(\l )}.}
Since the L.H.S. is easily seen to be of the form
${\l^{2}\/ 4}+O(\l^{4})$ the integration constant is determined by requiring
the R.H.S. to be analytic at $\l =0$. Adding also the contribution from the
term $A$ in
\refbr{sewedvertex}, the logarithms cancel, leaving the $k^2$ term
\eq{-k^{2}{\pi \/ 2}{K(\sqrt{1-\l^{2}})\/ K(\l)}.}

Collecting the results of the sewing procedure, we arrive at the following
expression
for the twisted scalar four-vertex:
\eqs{
\hat{W}_{4}^{(tw)}=\left( {2\/ \pi}(1-\l^{2})^{1\/ 4}K(\l)
\right)^{-{d\/ 2}}\int dk
:\exp{\left(-k^{2}{\pi\/2}{K(\sqrt{1-\l^{2}})\/K(\l)}+
2k{1\/r}\left[\Yps{-r}\tilde{c}^{(1)}_r+\Yps{r}\tilde{c}^{(2)}_r
\right]\right.} \lb
+\left.{1\/2}{1\/{rs}}\left[(\Ytn{-r}{-s}-
\Xtn{-r}{-s})\hat{c}^{(1)}_r\hat{c}^{(1)}_s -
2(\Ytn{-r}{s}+\Xtn{-r}{s})\hat{c}^{(1)}_r\hat{c}^{(2)}_s +
(\Ytn{r}{s}-\Xtn{r}{s})\hat{c}^{(2)}_r\hat{c}^{(2)}_s\right]\right):. }
The coefficients for some of the $\hat{c}_{r}^{(i)}\hat{c}_{s}^{(j)}$ and
$k\hat{c}_{r}^{(i)}$ terms
can be found in tables~\ref{cctabell} and~\ref{kctble} respectively.
\begin{table} \eqnono{\begin{array}{|c|c|c|c|} \hline
r,s & \hat{c}_{r}^{(1)}\hat{c}_{s}^{(1)} & \hat{c}_{r}^{(1)}\hat{c}_{s}^{(2)}
& \hat{c}_{r}^{(2)}\hat{c}_{s}^{(2)} \\ \hline
{1 \/ 2},{1\/ 2} & -{\l \/ 2}+{\linv}\left( 1-{E\/ K}\right)  &
-2(1-\l^{2})^{-{1\/ 2}}\left(1-\l^{2}-{E\/ K}\right) &
\l (1-\l^{2})^{-1}\left(1-{\l^{2}\/ 2}-{E\/ K}\right) \\ \hline
-{1\/ 2},{1\/ 2} & {E\/ K}(1-\l^{2})^{-{1\/ 2}}-1 & 2\l (1-\l^{2})^{-1}\left(
1-{E\/ K}\right) & {E\/ K}(1-\l^{2})^{-{1\/ 2}}-1 \\ \hline
{1\/ 2},-{1\/ 2} & {E\/ K}(1-\l^{2})^{-{1\/ 2}}-1  &
{2\/ \l}\left(1-{E\/ K}\right) &  {E\/ K}(1-\l^{2})^{-{1\/ 2}}-1
 \\ \hline
{3\/ 2},{1\/ 2} & {\l^{-2}\/ 3}\left(1-\l^{2}+{3\l^{4}\/ 8}-\right. &
-{2\/ 3\l}(1-\l^{2})^{-{1\/ 2}}\left(
1-{5\l^{2}\/ 2} + \right.  &
{\l^{2}\/ 3}(1-\l^{2})^{-2}\left( 1-\l^{2}+{3\l^{4}\/ 8}-\right. \\
 & \left. {E\/ K}
(1-{\l^{2}\/ 2})\right) & \left. {3\l^{4}\/ 2}-{E\/ K}(1-{\l^{2}\/ 2})\right) &
\left. {E\/ K}
(1-{\l^{2}\/ 2})\right)\\ \hline
{3\/ 2},-{1\/ 2} & -{1\/ 3\l}(1-\l^{2})^{-{1\/ 2}}\left( 1-\l^{2}-\right.
 & {2\l^{-2}\/ 3}\left( 1+{\l^{2}\/ 2}-{E\/ K}
(1-{\l^{2}\/ 2})\right) & -{\l \/ 3}(1-\l^{2})^{-{3\/ 2}}\left(
1-\l^{2}-\right. \\
 & \left. {E\/ K}(1-{\l^{2}\/ 2})\right) & & \left. {E\/ K}(1-{\l^{2}\/
2})\right) \\ \hline
{3 \/ 2},{3\/ 2} & {\l^{-3}\/ 9}\left( 1-{4\l^{2}\/ 3}+{5\l^{4}\/ 4}-
{3\l^{6}\/ 8}-\right. & -{2\/ 9}(1-\l^{2})^{-{3\/ 2}}\left( 1-6\l^{2} +
9\l^{4}-\right. & {\l^{3}\/ 9}(1-\l^{2})^{-3}\left( 1-{3\l \/ 2}+{5\l^{4}\/ 4}
-\right. \\
& \left. {E\/ K}(1-\l^{2}+{\l^{4}\/ 4})\right) & \left. {21\l^{6}\/ 4}-
{E\/ K}(1-\l^{2}+{\l^{4}\/ 4})\right) & \left. {3\l^{6}\/ 8}-{E\/ K}
(1-\l^{2}+{\l^{4}\/ 4})\right) \\ \hline
{3\/ 2},-{3\/ 2} & -{1\/ 3}+{2\/ 9}(1-\l^{2})^{-{3\/ 2}}\left(
1-{3\l^{2}\/ 2}+\right. & {2\l^{-3}\/ 9}\left( 1+3\l^{2}+
{5\l^{4}\/ 4}-\right. & -{1\/ 3}+{2\/ 9}(1-\l^{2})^{-{3\/ 2}}\left(
1-{3\l^{2}\/ 2}+\right.\\
& \left. {\l^{4}\/ 2}+{E\/ 2K}(1-\l^{2}+{\l^{4}\/ 4})\right) &
\left. {E\/ K}(1-\l^{2}+{\l^{4}\/ 4})\right) &
\left. {\l^{4}\/ 2}+{E\/ 2K}(1-\l^{2}+{\l^{4}\/ 4})\right)
\\ \hline
{1\/ 2},{3\/ 2} & {\l^{-2}\/ 3}\left( 1+{3\l^{4}\/ 8}-\l^{2}-\right. &
-{2\l \/ 3}(1-\l^{2})^{-{3\/ 2}}\left( 1-{5\l^{2}\/ 2}+\right. &
{\l^{2}\/ 3}(1-\l^{2})^{-2}\left( 1-\l^{2}+\right. \\
& \left. {E\/ K}(1-{\l^{2}\/ 2})\right) & \left. {3\l^{4}\/ 2}-
{E\/ K}(1-{\l^{2}\/ 2})\right) & \left. {3\l^{4}\/ 8}-{E\/ K}
(1-{\l^{2}\/ 2})\right) \\ \hline
-{1\/ 2},{3\/ 2} & -{1\/ 3\l}(1-\l^{2})^{-{1\/ 2}}\left( 1-\l^{2}-\right. &
{2\l^{2}\/ 3}(1-\l^{2})^{-2}\left( 1+{\l^{2}\/ 2}-\right. &
-{\l \/ 3}(1-\l^{2})^{-{3\/ 2}}\left( 1-\l^{2}-\right. \\
& \left. {E\/ K}(1-{\l^{2}\/ 2})\right) & \left. {E\/ K}(1-{\l^{2}\/ 2})
\right) & \left. {E\/ K}(1-{\l^{2}\/ 2})\right) \\ \hline
\end{array}}
\caption{Coefficients of $\hat{c}\hat{c}$ in the exponent of
the vertex.}
\label{cctabell}
\end{table}

\begin{table}
\eqnono{\begin{array}{|c|c|c|} \hline
r & k\hat{c}_{r}^{(1)} & k\hat{c}_{r}^{(2)} \\ \hline
{1\/ 2} & -{\pi \/ K}\l^{-{1\/2}} & {\pi \/ K}\l^{1\/ 2}(1-\l^{2})^{-{1\/ 2}}
\\ \hline
-{1\/ 2} & {\pi \/ K}\l^{1\/ 2}(1-\l^{2})^{-{1\/ 2}} & -{\pi \/ K}\l^{-{1\/2}}
\\ \hline
{3 \/ 2} & -{\l^{-{3\/ 2}}\/ 3}{\pi \/ K}(1-{\l^{2}\/2}) &
{\l^{{3\/ 2}}\/ 3}{\pi \/ K}(1-{\l^{2}\/2})(1-\l^{2})^{-{3\/ 2}} \\ \hline
-{3\/ 2} &  {\l^{{3\/ 2}}\/ 3}{\pi \/ K}(1-{\l^{2}\/2})(1-\l^{2})^{-{3\/ 2}} &
-{\l^{-{3\/ 2}}\/ 3}{\pi \/ K}(1-{\l^{2}\/2}) \\ \hline
\end{array}}
\caption{Coefficients of $k\hat{c}$ in the exponent.}
\label{kctble}
\end{table}

\section{The Closed Form of the Vertex}

The results of the previous section were obtained by means of a modified
version
of the algorithm introduced in ref.~\cite{ENS2}.
In that paper a closed expression for the four-Ramond vertex was proposed and
shown to agree with the sewing result for a finite (but large) number of
external oscillator terms.
An analytic proof of the proposed closed expression was later given in
ref.~\cite{NS93}.
In this section we will propose a corresponding closed expression for the
vertex of
four twisted scalars, which by Taylor expansion can be shown to reproduce the
results  of
section \ref{Sewsec}.

Just like in the case of the Ramond string, the closed expression for the
four vertex has a very natural form dictated by the structure of the branch
cuts generated by the twisted fields, namely
\eqs{
\lefteqn{\hat{W}_{4}^{(tw)}=\left( {2\/ \pi}(1-\l^{2})^{1\/ 4}K(\l)
\right)^{-{d\/ 2}}\int dk:\exp{\left( -k^{2}{\pi \/ 2}{K(\sqrt{1-\l^{2}})\/
K(\l)}
\right. }} \lb
& &-\left.  k\oint_C\hat{x}^{(tw)}(V^{-1}(z))H(z)+\oint_C dz\oint_C
dw\hat{x}^{(tw)}(V^{-1}(z))G(z,w)
\hat{x}^{(tw)}(V^{-1}(w))\right):}
where the propagator $G(z,w)=G(V_1,V_2;z,w)$ is given by
\eqs{
\lefteqn{G(z,w)=(z-w)^{-2}\left( A(\l )\left[ \left( {V_{1}^{-1}(w)\/
V_{1}^{-1}(z)}{V_{2}^{-1}(w)\/ V_{2}^{-1}(z)}\right)^{1\/ 2}+
\left( {V_{1}^{-1}(z)\/
V_{1}^{-1}(w)}{V_{2}^{-1}(z)\/ V_{2}^{-1}(w)}\right)^{1\/ 2}\right]\right.} \lb
& &+\left. (1-A(\l))\left[ \left( {V_{1}^{-1}(z)\/
V_{1}^{-1}(w)}{V_{2}^{-1}(w)\/ V_{2}^{-1}(z)}\right)^{1\/ 2}+
\left( {V_{1}^{-1}(w)\/
V_{1}^{-1}(z)}{V_{2}^{-1}(z)\/ V_{2}^{-1}(w)}\right)^{1\/ 2}\right] \right),}
with
\eq{A(\l)=\l^{-2}(1-{E\/ K}),}
and where
\eq{H(z)=-{i\pi \/ 2\l zK(\l)}\left( {V_{1}^{-1}(z)\/
V_{2}^{-1}(z)}\right)^{1\/ 2}.}
Here
$\hat{x}^{(tw)}(V^{-1}(z))=\hat{x}^{(tw)}(V^{-1}_1(z))+
\hat{x}^{(tw)}_2(V^{-1}_2(z))$ and the
contour $C$ encloses the cuts.
When performing the integrals in the exponent we find that the square root
cuts in the propagator cancel against the cuts in the external fields leaving
only simple integrals around poles. By identifying the coefficients of the
external oscillators we again arrive at precisely the results given in tables
\ref{cctabell} and \ref{kctble}. In other words, the proposed closed expression
exactly reproduces the results of the sewing. Although this check does not
constitute a proof,  which should be possible to find
using the techniques of ref.~\cite{NS93}, it nevertheless provides a very
strong
argument in favour of the general validity of our expression.

\section{Conclusions}

In this paper we have demonstrated that string vertices with more than two
external
twisted scalar legs can be obtained by sewing together two twisted three
vertices \cite{SchJ1,CFa},
or rather, which is entirely equivalent in this case, two dual twisted scalar
vertices \cite{ENS1}.
This was done by applying a modified version of the algorithm that was
originally deviced for the same purpose
in the context of twisted fermions \cite{ENS2}. The main task of this algorithm
is
to systematize the calculation of higher-level terms in the sewn vertex that
appear after sewing as rather complicated expressions in terms of
infinite-dimensional
matrices. They have previously been computed only for the massless mode in the
fermionic case \cite{CO}-\cite{BCO}
and for the lowest lying mode in the scalar case \cite{SW2,SchJ2}. For the
twisted scalar case some additional results relevant
for the next to lowest level can be found in ref.~\cite{DGM2}. The algorithm
used here generalizes these
results to arbitrary levels. It would be interesting to derive the same results
analytically as was done in the Ramond
case in ref.~\cite{NS93}.

 It is likely that this algorithm can be applied directly also
to twisted scalars and fermions of arbitrary conformal dimension but again only
in connection with
four vertices. We do not carry out a separate discussion for vertices with
untwisted external legs
because adding such legs will not alter the nature of the technical problems we
are addressing here.
 For vertices with six or more twisted external legs (and no loops), however,
one encounters new problems which are not yet known how to handle
\cite{NSTunp}.
The same is true for sewing of twisted loops. Another direction of
generalization
is higher twists. Despite the fact that vertices of this kind turn out to be of
interest also in other circumstances, e.g. in the theory of codes \cite{Mon93},
very little work has been done
in this field. Some results were obtained for ${\bf Z}_3$ twisted fields  in
refs.~\cite{Mon93,EN93}
but a lot remains to be discovered.\\[10mm]
{\Large\bf Acknowledgement}\\[4mm]
We are grateful to P. Sundell for useful discussions at an early stage of this
work.

\end{document}